\documentclass{aa} 
\usepackage{natbib}
\usepackage{graphicx}
\usepackage{txfonts}

\usepackage[dvipsnames]{xcolor}
\usepackage{xcolor}
\definecolor{darkgoldenrod}{rgb}{0.72, 0.53, 0.04}

\usepackage[colorlinks = true,
          linkcolor = RoyalBlue,
      urlcolor = darkgoldenrod,
  citecolor = RoyalBlue,
anchorcolor = RoyalBlue, 
draft = false]{hyperref}

\begin{document} 

   \title{Machine learning technique for morphological classification \\ 
   of galaxies from SDSS. I. Photometry-based approach}

  \author{Vavilova I.B.$^{1}$, Dobrycheva D.V.$^{1}$, Vasylenko M.Yu.$^{1,2}$, Elyiv A.A.$^{1}$, Melnyk O.V. $^{1}$, Khramtsov V.$^{3}$}

   \institute{
$^{1}$Main Astronomical Observatory of the National Academy of Sciences of Ukraine, 27 Zabolotnogo St., Kyiv, 03143, Ukraine\\
$^{2}$Institute of Physics of the National Academy of Sciences of Ukraine, 46 avenu Nauki, Kyiv, 03028, Ukraine\\
$^{3}$Institute of Astronomy, V. N. Karazin Kharkiv National University, 35 Sumska St., Kharkiv, 61022, Ukraine\\}

\titlerunning{Photometry-based approach with machine learning to the morphological classification of the SDSS galaxies}
\authorrunning{Vavilova I.B., Dobrycheva D.V., Vasylenko M.Yu. et al.}

% \abstract{}{}{}{}{} 
% 5 {} token are mandatory
 
  \abstract
   {Machine Learning methods are effective tools in astronomical tasks for classifying objects by their individual features. One of the promising utility is related to the morphological classification of galaxies at different redshifts.}
   {1)  To test in details five supervised machine learning techniques to determine their performance and define the most effective among them for the automated morphological classification of the SDSS galaxies. 2) To test the influence of photometry, image, and spectral data on morphology classification. 3) To apply the best fitting machine learning methods for revealing the unknown morphological types of galaxies from the SDSS DR9 at $z<0.1$.}
   {We used different galaxy classification techniques: human labeling, multi-photometry diagrams, Naive Bayes, Logistic Regression, Support Vector Machine, Random Forest, k-Nearest Neighbors, and k-fold validation.}
   {We present results of a binary automated morphological classification of galaxies conducted by human labeling, multi-photometry, and supervised Machine Learning methods. We applied its to the sample of galaxies from the SDSS DR9 with redshifts of $0.02<z<0.1$ and absolute stellar magnitudes of $-24^{m}<M_{r}<-19.4^{m}$. To study the classifier, we used absolute magnitudes: $M_{u}$, $M_{g}$, $M_{r}$, $M_{i}$, $M_{z}$, $M_{u}-M_{r}$, $M_{g}-M_{i}$, $M_{u}-M_{g}$, $M_{r}-M_{z}$, and inverse concentration index to the center $R50/R90$.
   Using the Support vector machine classifier and the data on color indices, absolute magnitudes, inverse concentration index of galaxies with visual morphological types, we were able to classify 316 031  galaxies from the SDSS DR9 with unknown morphological types. }
   {The methods of Support Vector Machine and Random Forest with  Scikit-learn machine learning in Python provide the highest accuracy for the binary galaxy morphological classification: 96.4\% correctly classified (96.1\% early $E$ and 96.9\% late $L$ types) and 95.5\% correctly classified (96.7\% early $E$ and 92.8\% late $L$ types), respectively. Applying the Support Vector Machine for the sample of 316 031 galaxies from the SDSS DR9 at $z<0.1$, we found 141 211 $E$ and 174 820 $L$ types among them.}
\keywords{Galaxies, galaxy morphology, machine learning methods}

\titlerunning{Machine learning for the automated galaxy morphological classification}
\authorrunning{Vavilova, Dobrycheva, Vasylenko et al.} 
 \maketitle 

\section{Introduction}
During the 1990s, the Artificial Neural Network (ANN) algorithms were intended for automatic morphological classification of galaxies since the huge extragalactic data sets have been conducted. A classification accuracy (or a success rate) of the ANNs was from 65\,\% to 90\,\% depending on the mathematical subtleties of the applied methods and the quality of galaxy samples. One of the first such works was made by \cite{1992AAS...181.6508S} with a feed-forward neural network, which has dealt with the classification of 5217 galaxies onto five classes (E, SO, Sa-Sb, Sc-Sd, and Irr) with a 64 \% accuracy. A detailed comparison of human and neural classifiers was presented by \cite{1995MNRAS.275..567N}, who used a Principal Component Analysis to classify 831 galaxies: the best result was with rms deviation of 1.8 T-types. Summarizing such first attempts, \cite{1995ApL&C..31...73L}, \cite{1996MNRAS.283..207L} resumed that ``the ANNs can replicate the classification by a human expert almost to the same degree of agreement as that between two human experts, to within 2~T-type units''.

An excellent introduction to the classification algorithms for astronomical tasks, including the morphological classification of galaxies, is given by \cite{2010IJMPD..19.1049B}, \cite{2012amld.book.....W}, \cite{2012cidu.conf...47V}, \cite{2014AAS...22325301V}, \cite{2015arXiv150305296A}, \cite{2020WDMKD..10.1349F}, \cite{2020kdbd.book..225E}, \cite{2020kdbd.book..307V} as well as, see, a classical work by \cite{2011arXiv1102.0550B} and a good pedagogical review with a discussion of the major methods, in which galaxies are studied morphologically and structurally, by \cite{2014MNRAS.444.1125C}.

The Sloan Digital Sky Survey (SDSS), which started in 2000, has collected more data in its first few weeks than had been amassed in astronomy history. Now, 20 years later, its archive contains about 170 terabytes of information. Soon its successor, the Large Synoptic Survey Telescope, will acquire that quantity of data every five days \citep{2000AJ....120.1579Y}. This provided entry points for the computer scientists wanting to get engaged in astronomical research and explains why big data mining and machine learning methods are gaining, so popularity to categorize celestial bodies in big datasets with more accuracy than ever. 

In this context, we review several works below, where different approaches were developed, and great efforts were made to identify the morphological types of galaxies from the SDSS in the visual and in the automated modes.

\cite{2004MNRAS.348.1038B} have tested the Supervised ANN for morphological classifications and obtained that it may be applied without human intervention for the SDSS galaxies (correlations between predicted and actual properties were around 0.9 with rms errors of order 10\,\%). \cite{2004MNRAS.349...87D} developed a method that combines two machine learning algorithms: Locally Weighted Regression and ANN. They tested it with 310 images of galaxies from the NGC catalogue and obtained accuracy 95.11\,\%, and 90.36\,\%, respectively. Siddhartha et al. (2007) explored Support
Vector Machines, Random Forests, and Naive Bayes as the galaxy image classifiers and Principal Component
Analysis for the direct image pixel data compressing. They resulted in favor of the Random Forest method, but have cited the opinion of several astronomers on the successful perspective of galaxy classification by morphic features as: ``one of the most cumbersome areas in celestial classification, and the one that has proven the most difficult to automate''.
Nevertheless, \cite{2010A&A...522A..21A} applied a probabilistic classification algorithm to classify the SDSS bright galaxies and obtained that it produces reasonable morphological classes and object-to-class assignments without any prior assumptions. 

\cite{2019MNRAS.490.2367C} in their work with a machine learning approach for the prediction of galaxies' dark matter halo masses used \textit{XGBoost}, Random Forest, and neural network. The sets of synthetic galaxy catalogs were used as training samples to built by populating dark matter halos in N-body simulations. Matching both the clustering and the joint-distributions of galaxy properties they were able to obtain halo masses of galaxies from the SDSS DR7 sample.

As for the visual morphological classification conducted during the last years, we note as follows. \cite{2010ApJS..186..427N} prepared the detailed visual classifications for 14\,034 galaxies from the SDSS\,DR4 at $z<0.1$, which can be used as a good training sample for calibrating the automated galaxy classification algorithms. A significant study was conducted by \cite{2010MNRAS.406..342B}, where galaxies from the Galaxy Zoo  Project\footnote{http://data.galaxyzoo.org} have formed a training sample for morphological classifications of galaxies from the SDSS\,DR6 into three classes (early types, spirals, spam objects). These authors convincingly showed that using a set of certain galaxy parameters, the Neural Network can reproduce the human classifications to better than 90\,\% for all these classes and that the Galaxy Zoo catalog (GZ1) can serve as a training sample. 

The hundreds of thousands of volunteers were involved into the Galaxy Zoo project to make visual classification of a million galaxies in the SDSS \citep{2008MNRAS.389.1179L}. Most of their results have found good scientific applications. For example, using the raw imaging data from SDSS that was available in the GZ1, and the handpicked galaxy's features from the SDSS, \cite{2012APS..APR.E1075K} applied a logistic regression classifier and attained 95.21\,\% classification accuracy. \cite{2013MNRAS.435.2835W} issued a new catalog of morphological types from the Galaxy Zoo Project (GZ2) in the synergy with the SDSS\,DR7, which contains more than 16 million morphological classifications of 304\,122 galaxies and their finer morphological features (bulges, bars, and the shapes of edge-on disks as well as parameters of the relative strengths of galactic bulges and spiral arms).

\cite{2016ApJS..223...20K} have generated a morphology catalog of the SDSS galaxies with the \textit{Wndchrm} image analysis utility using the nearest neighbor classifier. These authors pointed out that about 900\,000 of the instances classified as spirals and about 600\,000 of those classified as ellipticals have a statistical agreement rate of about 98\,\% with the Galaxy Zoo classification, see, also, \cite{2017MNRAS.464.4420S} on synergy of galaxy classification from CANDELS. Murrugarra et al. (2017) evaluated the Convolutional Neural Network to classify galaxies from the SDSS onto two classes as ellipticals/spirals by image processing and attained accuracy 90 -- 91\,\%. Using the same machine learning technique, the Convolutional Neural Network, especially inception method, Rahman et al. (2018) conducted classification into three general categories: ellipticals, spirals, and irregulars. They used 710 images (206 $E$, 320 $Sp$, 184 $Irr$) and obtained that images, which have went through image processing, showed a rather poor testing accuracy compared to not using any form of image processing. Their best testing accuracy was 78.3\,\%. Both supervised and unsupervised methods were applied by Gauthier et al. (2016) to study the Galaxy Zoo dataset of 61\,578 pre-classified galaxies (spiral, elliptical, round, disk). They found that the variation of galaxy images are correlated with brightness and eccentricity, the Random Forest method gives the best accuracy (67\,\%), meanwhile its combination with regression to predict the probabilities of galaxies associated with each class allows to reach a 94\,\% accuracy.

Wherein the photometric and spectral parameters of each object, as well as their images, are available through the SDSS website. It allows to use a well-known fact that galaxy morphological type is correlated with the color indices, luminosity, de Vaucouleurs radius, inverse concentration index, etc. In series of our work, we have demonstrated an effectiveness of a combination of the visual classification and the two-dimensional diagrams of color indices ${g-i}$ and one of the aforementioned parameters \citep{2009AN....330.1004V, 2012Ap.....55..293M, 2012AASP....2...42D, 2015KosNT..21c..94V}, see, also, for human and machine intelligence in GZoo projects \citep{2018MNRAS.476.5516B}. Namely, using the ``color indices vs inverse concentration indexes'' diagrams for each galaxy with radial velocities $3000<V<9500$ km/s from the SDSS\,DR5 we obtained criteria for separating the galaxies into three classes, specifically: (${E}$) early types -- elliptical and lenticular, (${S}$) spiral $Sa-Scd$, and (${LS}$) late spiral $Sd-Sdm$ and irregular $Im/BCG$ galaxies. Making a ternary automatic morphological galaxy classification (Fig. \ref{fig1}) we attained a good accuracy 98\,\% for ${E}$, 88\,\% for ${S}$, and only 57\,\% for ${LS}$ types. This approach based on the photometric data only (multi-parametric diagrams) was applied by us to classify a sample of 316 031 SDSS galaxies at $0.003\leq z \leq0.1$ from the SDSS\,DR9 (142\,979 $E$, 112\,240 $S$, 60\,812 $L$ type \citep{2017PhDT.......107D, 2013OAP....26..187D}\footnote{http://leda.univ-lyon1.fr/fG.cgi?n=hlstatistics\&a=htab\&z=d\&sql=iref= =52204}). The more detailed explanation is given by \cite{2018KPCB...34..290D}. 

This work deals with the automated morphological classification of the low redshift galaxies from the SDSS\,DR9 \& DR16. We used the cosmological WMAP7 parameters $\Omega_{M} = 0.27$, $\Omega_{\Lambda} = 0.73$, $\Omega_{k} = 0$, $ H_{0} = 0.71 $ and put the following tasks:

\begin{itemize}
    \item to verify various machine learning methods for selection of more effective among them to classify galaxies at $z<0.1$  with unknown morphological types from the SDSS\,DR9;
    \item to determine margins where the automated morphology classification based on the photometric parameters of galaxies gives the best result, including peculiarities of the Hubble/Vaucouleurs type of galaxy forms at different redshifts; to reveal typical problem points;
    \item to apply the developed criteria for the automatic morphological classification of galaxies at $z<0.1$ from the SDSS\,DR9 with unknown morphological types.
\end{itemize}

\section{Galaxy samples from the SDSS DR9 for the automated morphological classification }
\subsection{Galaxy sample} 
A preliminary sample of galaxies at $z<0.1$ with the absolute stellar magnitudes $-24^{m}<M_{r}<-13^{m}$ from the SDSS\,DR9 contained of $\sim$ 724\,000 galaxies. Following the SDSS recommendation, we input limits $m_{r} < 17.7$ by visual stellar magnitude in $r$-band to avoid typical statistical errors in spectroscopic flux. After excluding the duplicates of galaxy images and ``spam'' objects the final sample contained of N = 316\,031 galaxies. To clear the sample from duplicates of images of the same galaxy we used own code based on the minimal angle distances between such SDSS objects.

\begin{figure}
\centering
    \includegraphics[width=0.5\textwidth]{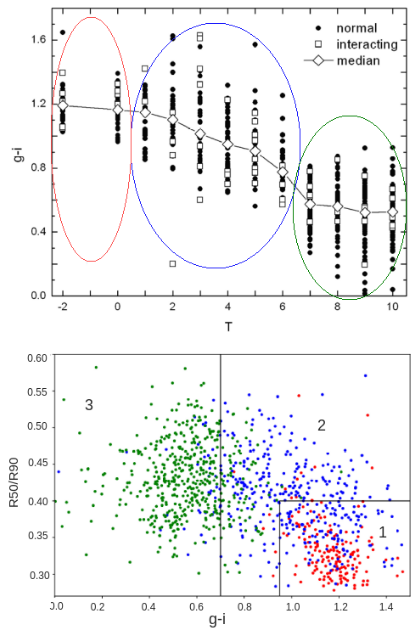}
    \caption{(Top) Dependence of the morphological types T on the color indices $\it{g--i}$ for 730 galaxies from the SDSS\,DR5. (Down) The inverse concentration indexes R50/R90 as functions of color indices for these galaxies; the red circles correspond to early types (-2--0), the blue circles to spirals (1-6), and the green circles to late type spiral and irregular galaxies. The lines define three regions into which a maximum number (> 90\,\%) of galaxies of morphological types (-2--0), (1--6), and (7--10) falls, respectively, with a minimum number of other morphological types}
    \label{fig1}
\end{figure}

\begin{figure}
	\includegraphics[width=0.5\textwidth]{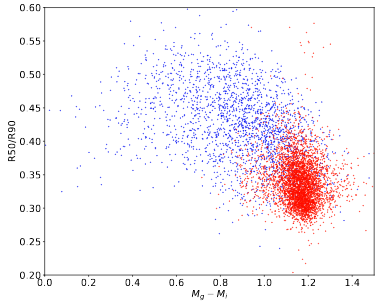}
    \caption{Diagram of color indices $g-i$ and inverse concentration indexes $R50/R90$ of training sample (6163 galaxies randomly selected with different redshifts and luminosities from the SDSS DR9).
		The red color indicates the visually classified galaxies (human labeling) of early $E-S0$ types, and the blue color indicates late $Sa-Irr$ types.}
    \label{fig2}
\end{figure}

\begin{figure}
	\includegraphics[width=0.5\textwidth]{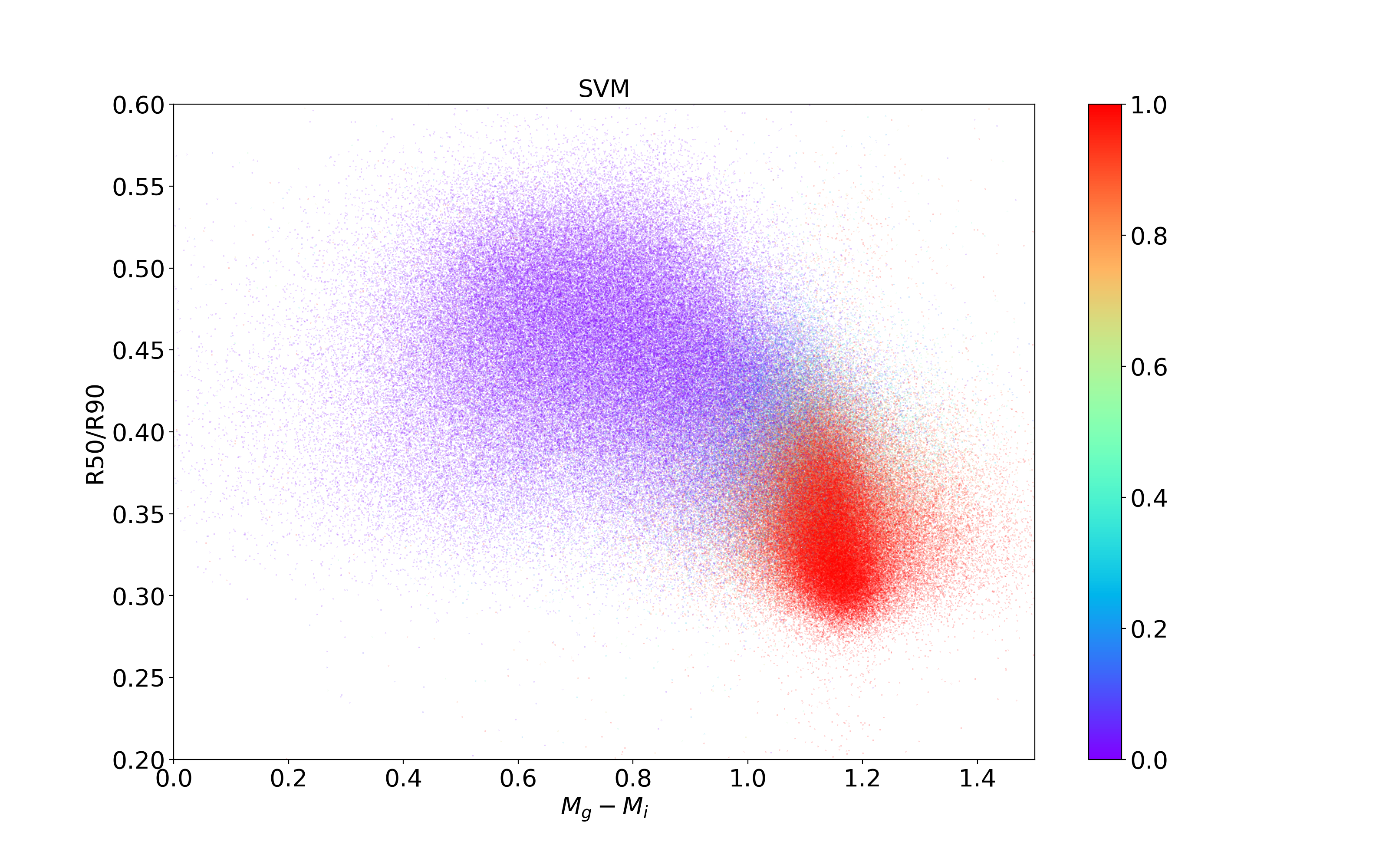} %\linewidth or \textwidth
    \caption{Diagram of color indices $g-i$ and inverse concentration indexes $R50/R90$ of 316\,031 galaxies at $z<0.1$ from the SDSS DR9 after applying the Support Vector Machine (SVM) method: red color -- early $E$ (from elliptical to lenticular) and blue color -- late $L$  (from $S0a$ to irregular $Im/BCG$) types. Color bar from 0 to 1 shows SVM probability to classify galaxy as the late to the early morphological type }
    \label{fig3}
\end{figure}

The absolute stellar magnitude of the galaxy was obtained by the formula 
$$
M_{r}=m_{r}-5\cdot\lg(D_{L})-25-K_{r}(z)-ext_{r},
$$
where $m_{r}$ - visual stellar magnitude in $\mathit r$ band, $D_{L}$ - luminosity distance, 
$ext_{r}$ - the Galactic absorption in $\mathit r$ band in accordance to \cite{1998ApJ...500..525S},  
$K_{r}(z)$ - k-correction in $\mathit r$ band according to \cite{2010MNRAS.405.1409C, 2012MNRAS.419.1727C}. 

The color indices were calculated as 
$$M_{g}-M_{i} = (m_{g}-m_{i}) - (ext_{g} - ext_{i}) - (K_{g}(z) - K_{i}(z)),$$
where $m_{g}$ and $m_{i}$ - visual stellar magnitude in $\mathit g$ and $\mathit i$ band;
$ext_{g}$ and $ext_{i}$ - the Galactic absorption in $\mathit g$ and $\mathit i$ band;
$K_{g}(z)$ and $K_{i}(z)$ - k-correction in $\mathit g$ and $\mathit i$ band, respectively. 

A ternary morphological classification with the method of multi-parametric diagrams (in-box classification) do not attain a reasonable accuracy to classify spiral galaxies of $Sa-Scd$ type (see, Section 1 and Fig. \ref{fig1} as well as \cite{2015Ap.....58..168D, 2018KPCB...34..290D, 2020kdbd.book..307V}). 

So, for verifying various supervised machine learning methods we decided to provide a binary automated morphological classification: early type galaxies $E$ -- from ellipticals to lenticulars; late type galaxies $L$ -- from $S0a-Sdm$ to irregular $Im/BCG$ galaxies. 

\subsection{Training samples} 
The supervised machine learning methods are searching for relationship between the input and output data, in our case, between features of galaxies (photometric parameters) and their morphological types. A training sample should represent as more as possible of these features allowing to generalize and to build the model for the prediction of the target variables in an unlabeled test sample (see, for example, \cite{2017arXiv170404650K}. That is why our first step before applying the machine learning methods was to compose a good training sample. 

We identified visually the morphological types ($E$ and $L$ ) of 6\,163 galaxies from the sample described in Section 2.1, which were randomly selected at different redshifts and with different luminosity. This is $\sim 2 \%$ of a total number of the studied galaxy sample (see, Section 4 for discussion and Fig.~\ref{fig4}).

Using one of the three color indices and such parameters as the inverse concentration index,
absolute stellar magnitude, de Vaucouleurs radius, and scale radius ($color-R50/R90$, $color-M_{r}$,  $color-deVRad_r$, and $color-expRad_r$ diagrams, respectively), it is possible to carry out a reliable preliminary morphological classification without invoking visual inspection. The dependence of the color indices and the parameter $R50/R90$ gives the best fitting because of the values of the parameters do not depend on the radial galaxy velocity and that the selection effects are avoid \citep{2012AASP....2...42D}. As the example, see a diagram of inverse concentration indexes $R50/R90$ as a function of color indices $g-i$ for 6\,163 galaxies of the training sample, which is shown in Fig.~\ref{fig2}. It demonstrates a good separation onto the early and late galaxy types and reveals also clearly the well-known bimodality color indices effect \citep{2004ApJ...615L.101B, 2014MNRAS.440..889S}. The overlap of the types in range of $M_{g}-M_{i}$ from 1.1 to 1.3 is still substantial and will be discussed in Section 5.

\textbf{We used the classification criterion for the color index $g - i$ and $R50/R90$. In this case, the criterion was determined visually by the graph of the relationship between these two values. The accuracy of the method for E types was 96\%, and for L types - 67\%, however, the training sample at that time we had a test, which means that the actual accuracy was at least a few percent lower.}

\section{The supervised machine learning methods and morphological classification}
The learning can be supervised, semi-supervised, unsupervised, and reinforcement (Burkov, ``The Hundred-page Machine Learning Book'', 2019). In our work we used only the supervised methods, where the dataset is collection of the labeled examples ${(x_{i}, y_{i})}^{N}_{i=1}$. 

In our case, each element $x_{i}$ among N is a galaxy feature vector, in which each dimension $j = 1, ..., D$ contains a value that describes $y_{i}$. That value is called a feature and is denoted as $x^{(j)}$. For instance, if each example \textbf{x} in our collection represents a galaxy, then the first feature, $x^{(1)}$, could contain absolute magnitude $M_{u}$, the second feature, $x^{(2)}$, could contain color indices $M_{u} - M_{r}$, and $x^{(3)}$ could contain the inverse concentration index $R50/R90$. Summing up, there are absolute magnitudes $M_{u}, M_{g}, M_{r}, M_{i}, M_{z}$, color indices $M_{u} - M_{r}, M_{g} - M_{i}, M_{r} - M_{z}$, inverse concentration indexes $R50/R90$ to the center. For all examples in the dataset, the feature at position $j$ in the feature vector always contains the same kind of information. It means that if $x^{(2)}_{i}$ contains color indices $M_{u} - M_{r}$ for some example $x_{i}$, then $x^{(2)}_{k}$ will also contain color indices $M_{u} - M_{r}$ in each example $x_{k}$, $k = 1, ... , N$. The label $y_{i}$ can be either an element belonging to a finite set of classes ${1, 2, . . . , T}$, or a real number, or a more complex structure, like a vector, a matrix, a tree, or a graph. In our work we have only two classes, ${E, L}$, where $E$ means the early type of galaxy and $L$ means the late morphological type.

The goal of a supervised learning algorithm is to use the dataset for producing a model that takes a feature vector $x$ as input and output information allowing to deduce the label for this feature vector. For instance, the model with a dataset of galaxies could take a feature vector describing the morphological type of galaxy as the input information and a probability that the galaxy has $E$ or $L$ morphological type as the output information.

We applied software with an open-source KNIME Analytics Platform ver. 3.7.0 \footnote{https://www.knime.com/}, which is intended for prediction of data classification with different machine learning methods and is actively used in the data science. Using KNIME we built and trained following classifiers: Naive Bayes, Random Forest, Support Vector Machine based on WEKA software, and neutral networks (RProp MLP). Using Scikit-Learn machine learning library (ver. 0.2.2 for the Python programming language), which is a simple tool for data mining and data analysis (see, for example, Ivezic and Babu, Statistical Challenges in Astronomy, 2014), we trained Naive Bayes, Random Forest, Support Vector Machine, K-nearest Neighbor, and Logistic Regression. 

For training the classifier, we used the absolute magnitudes: $M_{u}, M_{g}, M_{r}, M_{i}, M_{z}$, color indices $M_{u} - M_{r}, M_{g} - M_{i}, M_{r} - M_{z}$, and inverse concentration index $R50/R90$ (Section 2.2). 

\subsection{Naive Bayes}
The Naive Bayes classifiers are based on the Bayes theorem and conditional independence of the features to calculate the probability of class $G$ (in our case it is a morphological type of galaxies) with a given feature vector (set of galaxy attributes) $X\,=\,(x_{1},\ldots,x_{i})$ 

$$
p(G|X)=\frac{p(G)p(X|G)}{p(X)}.
$$

If we accept the conditional independence assumption, instead of computing the class-conditional probability for each combination of $X$, we only have to estimate the conditional probability of each $x_{i}$, given $G$. To classify a test record, the Naive Bayes Classifier computes the posterior probability for each class $G$:

$$
p(G|X)=\frac{p(G)\prod_{i=1}^{n} p(x_{i}|G)}{p(X)}.
$$

\subsection{Random Forest} 
The Random Forest classifiers works as follows: the training sample contains of $N$ objects (for example, 749 galaxies with  morphological types identified visually, fig.\ref{fig1}) dimension of objects feature is $M$ ($M_{u}, M_{g}, M_{r}, M_{i}, M_{z}, M_{u} - M_{r}, M_{g} - M_{i}, M_{r} - M_{z}$, inverse concentration indexes R50/R90 to the center) and the parameter $m$ is given (as usually $m=\sqrt{M}$) as an incomplete number of traits for training. Then we built the committee tree, where the most common way is as follows:

\begin{itemize}
        \item to generate a random subsample with size $N$ likely in the training sample. (Thus, some object will hit in two or more times, and on average $N{\displaystyle(1-1/N)^{N}}$, 
        and approximately $N/e$ objects will not hit in at all);
        \item to construct the decision tree that classifies the objects of this subsample. The next node of the tree in the process of creating will use not all $M$ objects feature, but only $m$, which are randomly chosen;
        \item to develop the tree up to the complete exhaustion of the subsample.
       \end{itemize}

Classification of objects is conducted by voting: each tree of the committee classifies the object to one of the classes, and the class wins if it has the most significant number of trees voted (Breiman, ``Machine Learning'', 2001).

\subsection{Support Vector Machine}
We get a training dataset of $n$ points of the form ($\vec{x}_{1}, y_{1}),\ldots, (\vec{x}_{n}, y_{n}$), where the $y_{i}$ are either 1 or $-1$ (in our work it is $E$ or $L$ morphological type of the galaxy). Each point indicates to which the point $\vec{x}_{i}$ belongs (set of attributes of galaxies). Each $\vec{x}_{i}$ is a $p$-dimensional real vector. We should find the ``maximum-margin hyperplane'' that divides the group of points $\vec{x}_{i}$, for which $y_{i}=1$ from the group of points and for which $y_{i}=-1$ is defined in such a manner that the distance between the hyperplane and the nearest point $\vec{x}_{i}$ from either group is maximized. Any hyperplane can be written as the set of points $\vec{x}_{i}$ satisfying $\vec{w}_{i}\vec{x}_{i}-b=0$, where $\vec{w}_{i}$ is the normal vector (not necessarily normalized) to the hyperplane. This is much likely Hesse normal form, except that $\vec{w}_{i}$ is not necessarily a unit vector. The parameter $\frac{b}{\Vert\vec{w}\Vert}$ determines the offset of the hyperplane from the origin along the normal vector $\vec{w}$ (VanderPlas, ``Python Data Science Handbook: Essential Tools for Working with Data'', 2016).

\subsection{K-Nearest Neighbors}
The classifier based on K-nearest neighbors is an example of the most straightforward machine learning algorithm. It does not create class-dividing functions, but remembers the position of training sample objects in the hyperspace of features. The disadvantage of this method is that its productivity linearly depends on the size of the training sample, the dependence from metrics, and the difficulty in selecting statistical weight. To implement this method, it is enough to choose the number of neighbors - K, the distance metric, find the K nearest neighbors in this metric and assign to object the class of the largest number of his neighbors. This method can be used not only for binary classification. In this case, the neighbors can be assigned a statistical weight of $1/d$, where d is the distance in the hyperspace of features. This meter is also sensitive to normalization, as all features must make the same contribution to the distance estimation. Finding the number K is important because it allows to describe the model avoiding retraining and undertraining (Raschka, ``Python Machine Learning'', 2015).  Depending on the metric of space, the distance will be determined in different ways, for example, in Euclidean space:

$$d_{i,j}=\sqrt{ \sum \limits_{k}|x_{i}-x_{j}|^2}$$

\subsection{Logistic Regression}
In Logistic Regression, we can model a morphological type of galaxy $y_{i}$ as a linear function of $x_{i}$. However, with a binary $y_{i}$ this is not straightforward because $wx_{i} + b$ is a function that spans from minus infinity to plus infinity, while $y_{i}$ has only two possible values (Burkov (2019), Raschka (2015)). For binary morphological classification, we should define a negative label as 0 and the positive label as 1, and we would need to find a simple continuous function whose codomain is (0, 1). In such a case, if the value returned by the model for input \textbf{x} is closer to 0, then we assign a negative label to \textbf{x}; otherwise, the example is labeled as positive. One function that has such a property is the standard logistic function (also known as the sigmoid function):

$$
f(x)=\frac{1}{1+e^{-x}}.
$$

\begin{figure}
	\includegraphics[width=0.5\textwidth]{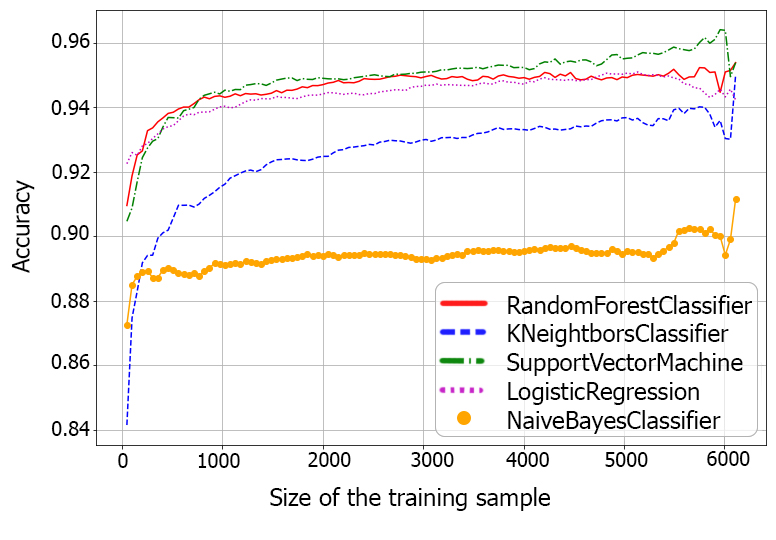}
    \caption{Verification of whether there are enough galaxies in a training sample to build a Machine Learning model. The green line (Support Vector Classifier), the red line (Random Forest), the rose line (Logistic Regression), the blue line (k-Nearest Neighbors), and the orange line (Naive Bayes) show the average accuracy of 10 repetitions of the evaluation procedure in Scikit-learn machine learning with Python.}
    \label{fig4}
\end{figure}

\section{Results}

We used the method of k-folds validation to estimate the accuracy with KNIME. Namely, we divided the sample into randomly selected five pieces, one by one, 4 of which served as a training and 1 as a test sample. Such a procedure was repeated five times and classification accuracy was defined as the average of the test samples. We set aside 20\% of the training sample to verify the accuracy of predicting of morphological types with Python. As a next step, we used the k-folds validation to predict the types in this delayed valid sample that were used to verify the accuracy of the method.

We have considered the accuracy change as a function of the sample size: if this function attains a level with the larger sizes, the existing set of training data is enough. However, if the accuracy continues to grow, most likely, it will not hurt to increase the amount of training data. To evaluate the accuracy of methods we have done the following procedures with a test sample of $N=6\,163$ galaxies for KNIME and Python software (Fig.~\ref{fig4}):

\begin{itemize}
    \item we divided training sample into subsamples changing the proportional sizes between train and valid samples.
    \item each of the subsample was formed 10 randomly times;
    \item then, these subsamples run with KNIME and Scikit-learn machine learning with Python for all the methods;
    \item then, an accuracy of methods was determined.
\end{itemize}

It turned out that the Support Vector Machine and the Random Forest classifiers (Fig.~\ref{fig4}) provide the highest accuracy of the automated binary galaxy morphological classification: 96.4\% correctly classified (96.1\% $E$ and 96.9\% $L$) and
95.5\% correctly classified (96.7\% $E$ and 92.8\% $L$),
respectively (Table \ref{tab1}). The attained accuracy of methods with KNIME is given in Table~\ref{tab2}.

\begin{table}
\caption{Accuracy (in \%) of the supervised machine learning methods for the automated binary morphological classification of galaxies from the SDSS DR9 at $z<0.1$ (early $E$ and late $L$ morphological types) in Python}
\begin{tabular}{ l  l  l  l  l }
\textbf{Classifier vs. Accuracy} & \textbf{Total} & \textbf{E type} & \textbf{L type} & \textbf{Error}\\
Naive Bayes                     & 89.0 & 92.0 & 82.0 & $\pm 1.0$ \\
k-Nearest Neighbors             & 94.5 & 93.9 & 95.8 & $\pm 0.6$ \\
Logistic Regression             & 94.9 & 96.8 & 91.1 & $\pm 0.6$ \\
\textbf{Random Forest}          & 95.5 & 96.7 & 92.8 & $\pm 0.3$ \\
\textbf{Support Vector Machine} & 96.4 & 96.1 & 96.9 & $\pm 0.6$ \\
\label{tab1}
\end{tabular}
\end{table}

\begin{table}
\caption{Accuracy (in \%) of the supervised machine learning methods for the automated binary morphological classification of galaxies from the SDSS DR9 at $z<0.1$ (early $E$ and late $L$ morphological types) in KNIME}
\begin{tabular}{ l  l  l  l  l }
\textbf{Classifier vs. Accuracy} & \textbf{Total} & \textbf{E type} & \textbf{L type} & \textbf{Error}\\
Naive Bayes                     & 88.9 & 84.4 & 91.0 & $\pm 0.1$ \\
k-Nearest Neighbors             & 93.7 & 93.9 & 93.6 & $\pm 0.5$ \\
Logistic Regression             & 94.8 & 93.8 & 
95.2 & $\pm 0.1$ \\
\textbf{Random Forest}          & 95.0 & 94.0 & 95.4 & $\pm 0.1$ \\
\textbf{Support Vector Machine} & 95.3 & 94.7 & 95.6 & $\pm 0.1$ \\
\label{tab2}
\end{tabular}
\end{table}

\begin{figure}
	\includegraphics[width=0.5\textwidth]{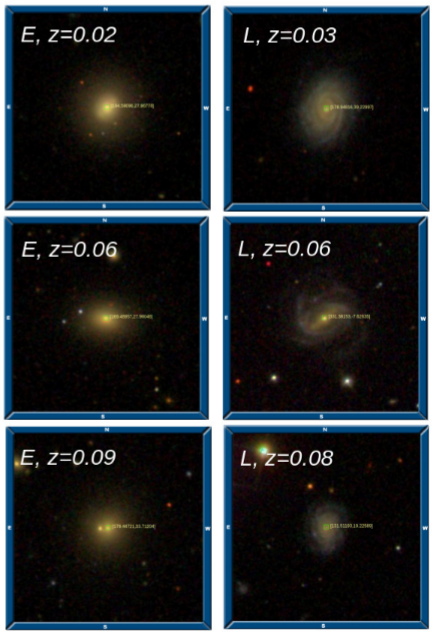}
    \caption{Examples of images of the SDSS galaxies at $z<0.1$ classified correctly as early $E$ and late $L$ morphological types}
    \label{fig5}
\end{figure}

So, using the data on the absolute stellar magnitudes, color indices, and inverse concentration indexes and coaching by Random Forest and Support Vector Machine classifiers to galaxies with visual morphological types, we applied these criteria to the studied sample of N~= 316\, 031 galaxies with unknown types and got their classification: 141\, 211 of early $E$ and 174\, 820 of late $L$ morphological types (Fig.~\ref{fig3}). The examples of galaxy images classified morphologically onto early and late types are given in Fig. \ref{fig5}.

\section{Discussion}
Various machine learning methods are helpful not only for the tasks of classification objects by morphological features of celestial bodies. They are effective for reconstruction of Zone of Avoidance \citep{2018RRPRA..23..244V}, finding gamma-ray sources for the upcoming Cherenkov Telescope Array \citep{2018AAS...23222003B}, spatio-temporal data \citep{2019arXiv190604928W}, classification of variable stars light curves \citep{2016A&A...587A..18K} and light-curve shape of a Type Ia supernova \citep{2020MNRAS.tmp.1851S}, determination of the distance modulus for local galaxies \citep{2020A&A...635A.124E}, prediction of galaxy halo masses \citep{2019MNRAS.490.2367C}, gravitational lenses search \citep{2019A&A...632A..56K}, automating discovery and classification of variable stars \citep{2012PASP..124.1175B} as well as for analyzing huge observational surveys, for example, the Zwicky Transient Facility \citep{2019PASP..131c8002M} or finding planets and exocomets from Kepler and TESS surveys \citep{2018nova.pres.4341K}. Besides traditional approach for classifying the galaxy types automatically in optical range, the machine learning methods demonstrate also a strong utility for classifying the radio galaxies types and peculiarities (\cite{2017ApJS..230...20A, 2018MNRAS.478.5547A, 2019EPSC...13..751W, 2019MNRAS.487.1729L, 2019PASP..131j8011R}). See, a very good review of machine lerning methods in astronomy in paper by \cite{2020WDMKD..10.1349F}.
Implementing machine learning methods for such astronomical tasks it's very useful to discuss their advantages and problem points, data quality regularity, and flexibility of classification pipeline. 

\subsection{Several problem points of the supervised machine learning methods for the automated morphological classification of galaxies from the SDSS}

The main problems of machine learning related to the morphological classification can be divided into two categories. The first one is related with a sample preparation, which includes: determining the parameters which are the best for dividing objects to classes, selecting a homogeneous dataset for classification parameters, creating a sub-directory for training algorithms, cleaning the sub-list of ''undesired'' (misclassified) objects, determining the best methods for the task, and selecting the best machine learning features to build training sample. 
The second category includes problems related to the individual peculiarities of selected objects and to the quality of image/photometry/spectrum galaxy data.

\textit{Selection of the best parameters of machine learning for training}. To determine the training parameters, we need the relationship between the accuracy of the model in training and test samples. In other words, we need to determine the parameters for which a) the accuracy on the test data will be maximized, b) the difference between the test and training data is minimal, and c) the accuracy on the training data is less than 100 \% to avoid of over fitting (see, for example, Fig. \ref{fig4}). The complexity is that these points do not always lie one below the other, So, the average values of the precision ratios on the training and test data should be analyzed for a large number of cycles to determine the best parameters \citep{2019OAP....32...46V}. 

\begin{figure}
	\includegraphics[width=0.5\textwidth]{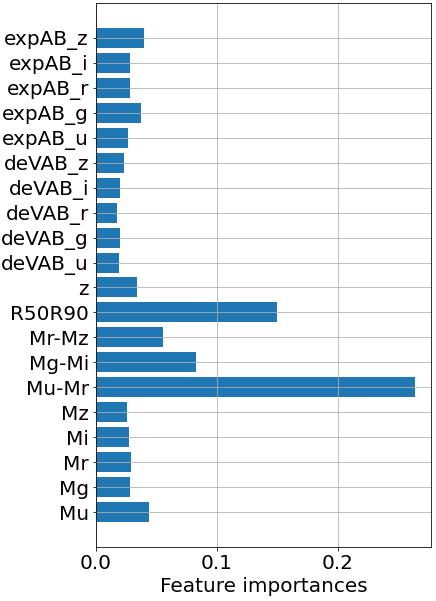}
    \caption{Estimation of the relative importance of some photometric parameters for the automated morphological classification of galaxies from the SDSS. Training sample contains of 11 000 galaxies. Parameters with higher values correspond to more significant ones. $deVAB_u$, $deVAB_g$, $deVAB_r$, $deVAB_i$, $deVAB_z$ - de Vaucouleurs radius fit $b/a$ in different bands, $expAB_u$, $expAB_g$, $expAB_r$, $expAB_i$, $expAB_z$ - exponential fit $b/a$ in different bands,  $z$ - redshift, $Mu-Mr$,$Mg-Mi$,$Mr-Mz$ - color indices, $Mu$,$Mg$,$Mr$,$Mi$,$Mz$ - absolute magnitudes, and $R50/R90$ - inverse concentration index to the centre}
    \label{fig6}
\end{figure}

\textit{Features of galaxies, which are the best fitted for the morphology classification onto types}. To determine such photometry parameters, we need a) to create a small homogeneous galaxy sample for training, where all the datasets have certain types available in the database; b) to test these parameters using, for example, the Fisher method for evaluating the significance of these features; c) to determine the distribution of galaxies of different types at different redshifts selecting sets of benchmarks by analyzing "slices" for one or several parameters (Fig. \ref{fig6}). 

\textit{Human labeling, automated classification, and image/ spectrum quality of the data.} We underline the problem points of the SDSS galaxy data, which led to their morphology misclassification, as follows: 1) interacting galaxies, 2) background galaxy, 3) stars covering the galaxy image, 4) artifacts (diffraction, satellites), 5) spiral galaxies of the red color, 6) bright nucleus (spiral galaxies are defined as ellipticals), 7) bad background, 8)	dim objects (low signal-to-noise ratio), 9) face-on and edge-on galaxies, 10) ''false'' objects (gravitational lenses etc.). 

\begin{figure}
	\includegraphics[width=0.5\textwidth]{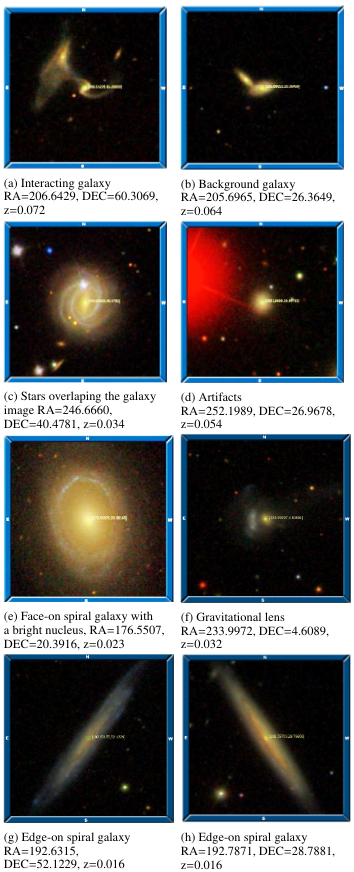}
    \caption{Typical examples of images of the SDSS DR16 galaxies at $z<0.1$ with the misclassified morphological types}
    \label{fig7}
\end{figure}

The examples of such objects are presented in Fig. \ref{fig7}. Because of they can be identified simply at the step of training sample building (see, subsection 2.2), we deleted such objects from training sample. Some objects may have incomplete data for comparison, including nonstandard spectra to compare its with the images for clarification of morphological types (for example, the galaxy with RA = 151.86, DEC = 18.85, z = 0.06854, Plate = 2372). 

Nevertheless, we may conclude that such misclassified morphologically objects contribute $\sim 1 \%$ error in the classification of a general galaxy sample (see, also, \cite{2018AJ....156..284C}).

\textit{Edge-on galaxies.} We used the Revised Flat Galaxy Catalogue, RFGC, 4444 galaxies \citep{1999BSAO...47....5K}, and Two-Micron Flat Galaxy Catalogue, 2MFGC, 18020 galaxies \citep{2015AstBu..70...24M} for cross-verification of the edge-on galaxies from our sample (which should be recognized as the late type spirals and never as the ellipticals) and for analysis of a contribution of this type error into the accuracy of the applied machine learning methods. The RFGC gives the data on coordinates, axis ratio, position angles, and names of 4444 galaxies including its in the Principal Galaxy Catalogue \citep{1989A&AS...80..299P}, but do not contain of information on the radial velocities or redshifts. Our sample contains of 934 flat galaxies from the RFGC as well as 3143 galaxies from the 2MFGC.   

We estimated prediction of morphological type in dependence on redshift by five machine learning technique and human labeling method. As you see from Table \ref{tab3}, Random Forest and Logistic Regression (Python) give the higher mean accuracy for all the range of redshifts. Nevertheless, our calculations have demonstrated that Naive Bayes give the highest accuracy 95 \% for $z \leq 0.02$. As for the prediction of edge-on galaxies as the late morphological type more preciously, we obtained accuracy 86 \% for Naive Bayes and 72 \% for Logistic Regression at $z \leq 0.05$ and $\sim 50 \% $ for Random Forest and Support Vector Machine at $z > 0.05$. As a result for these samples (galaxies from RFGC and 2MFGC) we may conclude that all five machine learning techniques and human labeling method provide determination of the correct edge-on type for 2/3 of total samples of galaxies.   

\begin{table}
\caption{Accuracy of determining spiral flat galaxies from  RFGC catalog}
\begin{tabular}{ l  l  l }
\textbf{Classifier} &  \textbf{Accuracy, \%}  \\
Human labeling               & 54.4   & \\
 &  \textbf{Python, \%} & \textbf{KNIME, \%} \\
Naive Bayes             & 70.2 & 73.7 \\
k-Nearest Neighbors     & 61.6 & 63.3 \\
Logistic Regression     & 71.6 & 63.6 \\
Random Forest           & 77.2 & 77.1 \\
Support Vector Machine  & 63.3 & 72.3 \\
\label{tab3}
\end{tabular}
\end{table}

At the same time, the results of applying the Deep convolutional neural network to the images of this sample \citep{2019OAP....32...21K} with the same aim of a binary
morphological classification has been shown that Deep Learning methods can classify rounded sources as ellipticals but it can not catch the spectral energy distribution properties of galaxies more clearly than Support Vector Machine trained on the photometric features of galaxies.

Generally, we have overestimated number of elliptical galaxies and underestimated number of spiral galaxies, when the face-on and edge-on galaxies are classified morphologically. This problem can be decided, when we form training samples through several steps (pre-training, fine-tuning, and classification). The step of fine-tuning includes the limitations on the axes-ratio for elliptical galaxies, additional photometry parameters for the face-on spiral galaxies as well as trainings with images and spectral features of galaxies (Khramtsov, 2020). 

This type of error is the same when we make a decision how to recognize morphological types of galaxies in region, where their photometry parameters are overlapping (see, Fig. \ref{fig2} for training and Fig. \ref{fig3} for the studied samples in region of $M_{g}-M_{i}$ from 1.1 to 1.3). Should we select only well-defined morphologically objects to avoid visual classification errors or add objects that are classified difficultly in the hope of finding more subtle features for each morphological type? To answer these questions, we added $\sim 5000$ galaxies from this region into training sample, which were selected by probability of determination 
of a certain type  $(50 \pm 5 \%)$. This allowed us to test accuracy changes and to define benefits from such a tuning. As you see from Table \ref{tab4}, such an approach worsened the result of the classification. The training sample in this case becomes subjective, depends on the human labeling rather than on the parameters on which the machine is based in the classification

\begin{table}
\caption{Accuracy (in \%) of the supervised machine learning methods for the automated binary morphological classification of 
modified  sample of 11301 galaxies from the SDSS DR9 at $z<0.1$ (early $E$ and late $L$ morphological types) in Python}
\begin{tabular}{ l  l  l  l  l }
\textbf{Classifier vs. Accuracy} & \textbf{Total} & \textbf{E type} & \textbf{L type} & \textbf{Error} \\
Naive Bayes                     & 66.8 & 64.1 & 70.4 & $\pm 1.2$ \\
k-Nearest Neighbors             & 79.4 & 80.3 & 78.6 & $\pm 0.7$ \\
Logistic Regression             & 81.9 & 83.9 & 80.3 & $\pm 0.3$ \\
\textbf{Random Forest}          & 82.4 & 87.6 & 78.6 & $\pm 0.4$ \\
\textbf{Support Vector Machine} & 84.3 & 89.0 & 80.6 & $\pm 0.5$ \\ %all table's rows have to be closed by \\, not &\ This symbol & creates one more column in the table!
\label{tab4}
\end{tabular}
\end{table}

\section{Conclusions}
We present the results of the automated morphological classification of galaxies from the SDSS with redshifts of $0.02<z<0.1$ and absolute stellar magnitudes of $-24^{m}<M_{r}<-19.4^{m}$.

Using the visual classification of galaxies and multi-parametric diagrams color-$M_{r}$, color-$R50/R90$, color-$deVRadr$, and color-$expRadr$, we discovered possible criteria for separating the galaxies into three classes: 1) early types, elliptical and lenticular; 2) spiral $Sa-Scd$, and 3) late spiral $Sd-Sdm$ and irregular $Im/BCG$ types. Due to a low accuracy for the $Sa-Scd$ types of galaxies, we concentrated our exploration on the automated classification onto two classes, $E$ early and $L$ late types of galaxies. 

We evaluated the accuracy of different supervised Machine learning methods to be applied to the binary automated morphological classification of galaxies (Naive Bayes, Random Forest, Support Vector Machines, Logistic Regression, and k-Nearest Neighbor algorithm). To study the classifier, we used absolute magnitudes: $M_{u}$, $M_{g}$, $M_{r}$, $M_{i}$, $M_{z}$, $M_{u}-M_{r}$, $M_{g}-M_{i}$, $M_{u}-Mg_{g}$, $M_{r}-M_{z}$, and inverse concentration index to the center $R50/R90$. 

We obtained that methods of Support Vector Machine and Random Forest with  Scikit-learn  machine  learning  in Python provide the highest accuracy for the binary galaxy morphological classification: 96.4\% correctly classified (96.1\% early $E$ and 96.9\% late $L$) types and 95.5\% correctly classified (96.7\% early $E$ and 92.8\% late $L$) types, respectively. It allowed us to create the Catalogue of morphological types of 316 031 galaxies from the SDSS at $z<0.1$ applying the Support Vector Machine, namely we revealed 141 211 $E$-type and 174 820 $L$-type galaxies among them. 

Analysis of problem points testifies that Support Vector Machine and Random Forest are the effective tools for the automated galaxy morphology classification.

\begin{acknowledgements}
We thank Prof. Massimo Capacciolli and Dr. Valentina Karachentseva for the helpful discussion and remarks. This work was partially supported in frame of the budgetary program of the National Academy of Sciences of Ukraine ”Support for the development of priority fields of scientific research” (CPCEL 6541230). 
This work was partially supported by the grant for Young Scientist’s Research Laboratories (2018-2019, Dobrycheva\,D.V.) and the Youth Scientific Project (2019-2020, Dobrycheva\,D.V., Vasylenko\,M.Yu.) of the National Academy of Sciences of Ukraine. HyperLeda \cite{2014A&A...570A..13M} and SDSS IV \cite{2017AJ....154...28B} were helpful to our study. This research has made use of the NASA/IPAC Extragalactic Database (NED), which is operated by the Jet Propulsion Laboratory, California Institute of Technology, under contract with the National Aeronautics and Space Administration.

\end{acknowledgements}

%-------------------------------------------------------------------

\bibliographystyle{aa} % style aa.bst
\bibliography{references} % your references Yourfile.bib

\begin{thebibliography}{61}
\expandafter\ifx\csname natexlab\endcsname\relax\def\natexlab#1{#1}\fi



\bibitem[Ahn \emph{et al.}(2012)]{2012ApJS..203...21A}Ahn, C.P., Alexandroff, R., Allende Prieto, C., Anderson, S.F., Anderton, T., Andrews, B.H., and, ...: 2012, {\it The Astrophysical Journal Supplement Series} {\bf 203}, 21. doi:10.1088/0067-0049/203/2/21.


\bibitem[Ahumada \emph{et al.}(2020)]{2020ApJS..249....3A}Ahumada, R., Allende Prieto, C., Almeida, A., Anders, F., Anderson, S.F., Andrews, B.H., and, ...: 2020, {\it The Astrophysical Journal Supplement Series} {\bf 249}, 3. doi:10.3847/1538-4365/ab929e.


\bibitem[Alger \emph{et al.}(2018)]{2018MNRAS.478.5547A}Alger, M.J., Banfield, J.K., Ong, C.S., Rudnick, L., Wong, O.I., Wolf, C., and, ...: 2018, {\it Monthly Notices of the Royal Astronomical Society} {\bf 478}, 5547. doi:10.1093/mnras/sty1308.


\bibitem[Al-Jarrah \emph{et al.}(2015)]{2015arXiv150305296A}Al-Jarrah, O.Y., Yoo, P.D., Muhaidat, S., Karagiannidis, G.K., and Taha, K.: 2015, {\it arXiv e-prints}, arXiv:1503.05296.


\bibitem[Andrae, Melchior, and Bartelmann(2010)]{2010A&A...522A..21A}Andrae, R., Melchior, P., and Bartelmann, M.: 2010, {\it Astronomy and Astrophysics} {\bf 522}, A21. doi:10.1051/0004-6361/201014169.

\bibitem[Aniyan and Thorat(2017)]{2017ApJS..230...20A}Aniyan, A.K. and Thorat, K.: 2017, {\it The Astrophysical Journal Supplement Series} {\bf 230}, 20. doi:10.3847/1538-4365/aa7333.


\bibitem[Ball \emph{et al.}(2004)]{2004MNRAS.348.1038B}Ball, N.M., Loveday, J., Fukugita, M., Nakamura, O., Okamura, S., Brinkmann, J., and, ...: 2004, {\it Monthly Notices of the Royal Astronomical Society} {\bf 348}, 1038. doi:10.1111/j.1365-2966.2004.07429.x.

\bibitem[Ball and Brunner(2010)]{2010IJMPD..19.1049B}Ball, N.M. and Brunner, R.J.: 2010, {\it International Journal of Modern Physics D} {\bf 19}, 1049. doi:10.1142/S0218271810017160.


\bibitem[Balogh \emph{et al.}(2004)]{2004ApJ...615L.101B}Balogh, M.L., Baldry, I.K., Nichol, R., Miller, C., Bower, R., and Glazebrook, K.: 2004, {\it The Astrophysical Journal} {\bf 615}, L101. doi:10.1086/426079.


\bibitem[Banerji \emph{et al.}(2010)]{2010MNRAS.406..342B}Banerji, M., Lahav, O., Lintott, C.J., Abdalla, F.B., Schawinski, K., Bamford, S.P., and, ...: 2010, {\it Monthly Notices of the Royal Astronomical Society} {\bf 406}, 342. doi:10.1111/j.1365-2966.2010.16713.x.


\bibitem[Beck \emph{et al.}(2018)]{2018MNRAS.476.5516B}Beck, M.R., Scarlata, C., Fortson, L.F., Lintott, C.J., Simmons, B.D., Galloway, M.A., and, ...: 2018, {\it Monthly Notices of the Royal Astronomical Society} {\bf 476}, 5516. doi:10.1093/mnras/sty503.


\bibitem[Bieker(2018)]{2018AAS...23222003B}Bieker, J.: 2018, {\it American Astronomical Society Meeting Abstracts \#232}.


\bibitem[Blanton \emph{et al.}(2017)]{2017AJ....154...28B}Blanton, M.R., Bershady, M.A., Abolfathi, B., Albareti, F.D., Allende Prieto, C., Almeida, A., and, ...: 2017, {\it The Astronomical Journal} {\bf 154}, 28. doi:10.3847/1538-3881/aa7567.

\bibitem[Bloom \emph{et al.}(2012)]{2012PASP..124.1175B}Bloom, J.S., Richards, J.W., Nugent, P.E., Quimby, R.M., Kasliwal, M.M., Starr, D.L., and, ...: 2012, {\it Publications of the Astronomical Society of the Pacific} {\bf 124}, 1175. doi:10.1086/668468.


\bibitem[Buta(2011)]{2011arXiv1102.0550B}Buta, R.J.: 2011, {\it arXiv e-prints}, arXiv:1102.0550.


\bibitem[Cabrera-Vives, Miller, and Schneider(2018)]{2018AJ....156..284C}Cabrera-Vives, G., Miller, C.J., and Schneider, J.: 2018, {\it The Astronomical Journal} {\bf 156}, 284. doi:10.3847/1538-3881/aae9f4.

\bibitem[Calderon and Berlind(2019)]{2019MNRAS.490.2367C}Calderon, V.F. and Berlind, A.A.: 2019, {\it Monthly Notices of the Royal Astronomical Society} {\bf 490}, 2367. doi:10.1093/mnras/stz2775.

\bibitem[de la Calleja and Fuentes(2004)]{2004MNRAS.349...87D}de la Calleja, J. and Fuentes, O.: 2004, {\it Monthly Notices of the Royal Astronomical Society} {\bf 349}, 87. doi:10.1111/j.1365-2966.2004.07442.x.


\bibitem[Chilingarian, Melchior, and Zolotukhin(2010)]{2010MNRAS.405.1409C}Chilingarian, I.V., Melchior, A.-L., and Zolotukhin, I.Y.: 2010, {\it Monthly Notices of the Royal Astronomical Society} {\bf 405}, 1409. doi:10.1111/j.1365-2966.2010.16506.x.

\bibitem[Chilingarian and Zolotukhin(2012)]{2012MNRAS.419.1727C}Chilingarian, I.V. and Zolotukhin, I.Y.: 2012, {\it Monthly Notices of the Royal Astronomical Society} {\bf 419}, 1727. doi:10.1111/j.1365-2966.2011.19837.x.

\bibitem[Conselice \emph{et al.}(2014)]{2014MNRAS.444.1125C}Conselice, C.J., Bluck, A.F.L., Mortlock, A., Palamara, D., and Benson, A.J.: 2014, {\it Monthly Notices of the Royal Astronomical Society} {\bf 444}, 1125. doi:10.1093/mnras/stu1385.



\bibitem[Dobrycheva and Melnyk(2012)]{2012AASP....2...42D}Dobrycheva, D. and Melnyk, O.: 2012, {\it Advances in Astronomy and Space Physics} {\bf 2}, 42.


\bibitem[Dobrycheva(2013)]{2013OAP....26..187D}Dobrycheva, D.V.: 2013, {\it Odessa Astronomical Publications} {\bf 26}, 187.

\bibitem[Dobrycheva \emph{et al.}(2015)]{2015Ap.....58..168D}Dobrycheva, D.V., Melnyk, O.V., Vavilova, I.B., and Elyiv, A.A.: 2015, {\it Astrophysics} {\bf 58}, 168. doi:10.1007/s10511-015-9373-x.

\bibitem[Dobrycheva(2017)]{2017PhDT.......107D}Dobrycheva, D.V.: 2017, {\it Ph.D. Thesis}.

\bibitem[Dobrycheva \emph{et al.}(2018)]{2018KPCB...34..290D}Dobrycheva, D.V., Vavilova, I.B., Melnyk, O.V., and Elyiv, A.A.: 2018, {\it Kinematics and Physics of Celestial Bodies} {\bf 34}, 290. doi:10.3103/S0884591318060028.

\bibitem[El Bouchefry and de Souza(2020)]{2020kdbd.book..225E}El Bouchefry, K. and de Souza, R.S.: 2020, {\it Knowledge Discovery in Big Data from Astronomy and Earth Observation}, 225. doi:10.1016/B978-0-12-819154-5.00023-0.

\bibitem[Elyiv \emph{et al.}(2020)]{2020A&A...635A.124E}Elyiv, A.A., Melnyk, O.V., Vavilova, I.B., Dobrycheva, D.V., and Karachentseva, V.E.: 2020, {\it Astronomy and Astrophysics} {\bf 635}, A124. doi:10.1051/0004-6361/201936883.


\bibitem[Fluke and Jacobs(2020)]{2020WDMKD..10.1349F}Fluke, C.J. and Jacobs, C.: 2020, {\it WIREs Data Mining and Knowledge Discovery} {\bf 10}, e1349. doi:10.1002/widm.1349.

\bibitem[Karachentsev \emph{et al.}(1999)]{1999BSAO...47....5K}Karachentsev, I.D., Karachentseva, V.E., Kudrya, Y.N., Sharina, M.E., and Parnovskij, S.L.: 1999, {\it Bulletin of the Special Astrophysics Observatory} {\bf 47}, 5.

\bibitem[Kates-Harbeck(2012)]{2012APS..APR.E1075K}Kates-Harbeck, J.: 2012, {\it APS April Meeting Abstracts}.

\bibitem[Khramtsov \emph{et al.}(2019)]{2019OAP....32...21K}Khramtsov, V., Dobrycheva, D.V., Vasylenko, M.Y., and Akhmetov, V.S.: 2019, {\it Odessa Astronomical Publications} {\bf 32}, 21. doi:10.18524/1810-4215.2019.32.182092.

\bibitem[Khramtsov \emph{et al.}(2019)]{2019A&A...632A..56K}Khramtsov, V., Sergeyev, A., Spiniello, C., Tortora, C., Napolitano, N.R., Agnello, A., and, ...: 2019, {\it Astronomy and Astrophysics} {\bf 632}, A56. doi:10.1051/0004-6361/201936006.

\bibitem[Kim and Bailer-Jones(2016)]{2016A&A...587A..18K}Kim, D.-W. and Bailer-Jones, C.A.L.: 2016, {\it Astronomy and Astrophysics} {\bf 587}, A18. doi:10.1051/0004-6361/201527188.

\bibitem[Kohler(2018)]{2018nova.pres.4341K}Kohler, S.: 2018, {\it AAS Nova Highlights}, 4341.

\bibitem[Kremer \emph{et al.}(2017)]{2017arXiv170404650K}Kremer, J., Stensbo-Smidt, K., Gieseke, F., Steenstrup Pedersen, K., and Igel, C.: 2017, {\it arXiv e-prints}, arXiv:1704.04650.

\bibitem[Kuminski and Shamir(2016)]{2016ApJS..223...20K}Kuminski, E. and Shamir, L.: 2016, {\it The Astrophysical Journal Supplement Series} {\bf 223}, 20. doi:10.3847/0067-0049/223/2/20.

\bibitem[Lahav(1995)]{1995ApL&C..31...73L}Lahav, O.: 1995, {\it Astrophysical Letters and Communications} {\bf 31}, 73.


\bibitem[Lahav \emph{et al.}(1996)]{1996MNRAS.283..207L}Lahav, O., Naim, A., Sodr{\'e}, L., and Storrie-Lombardi, M.C.: 1996, {\it Monthly Notices of the Royal Astronomical Society} {\bf 283}, 207. doi:10.1093/mnras/283.1.207.

\bibitem[Lintott \emph{et al.}(2008)]{2008MNRAS.389.1179L}Lintott, C.J., Schawinski, K., Slosar, A., Land, K., Bamford, S., Thomas, D., and, ...: 2008, {\it Monthly Notices of the Royal Astronomical Society} {\bf 389}, 1179. doi:10.1111/j.1365-2966.2008.13689.x.

\bibitem[Lukic \emph{et al.}(2019)]{2019MNRAS.487.1729L}Lukic, V., Br{\"u}ggen, M., Mingo, B., Croston, J.H., Kasieczka, G., and Best, P.N.: 2019, {\it Monthly Notices of the Royal Astronomical Society} {\bf 487}, 1729. doi:10.1093/mnras/stz1289.

\bibitem[Mahabal \emph{et al.}(2019)]{2019PASP..131c8002M}Mahabal, A., Rebbapragada, U., Walters, R., Masci, F.J., Blagorodnova, N., van Roestel, J., and, ...: 2019, {\it Publications of the Astronomical Society of the Pacific} {\bf 131}, 038002. doi:10.1088/1538-3873/aaf3fa.

\bibitem[Makarov \emph{et al.}(2014)]{2014A&A...570A..13M}Makarov, D., Prugniel, P., Terekhova, N., Courtois, H., and Vauglin, I.: 2014, {\it Astronomy and Astrophysics} {\bf 570}, A13. doi:10.1051/0004-6361/201423496.

\bibitem[Melnyk, Dobrycheva, and Vavilova(2012)]{2012Ap.....55..293M}Melnyk, O.V., Dobrycheva, D.V., and Vavilova, I.B.: 2012, {\it Astrophysics} {\bf 55}, 293. doi:10.1007/s10511-012-9236-7.


\bibitem[Mitronova and Korotkova(2015)]{2015AstBu..70...24M}Mitronova, S.N. and Korotkova, G.G.: 2015, {\it Astrophysical Bulletin} {\bf 70}, 24. doi:10.1134/S1990341315010034.


\bibitem[Naim \emph{et al.}(1995)]{1995MNRAS.275..567N}Naim, A., Lahav, O., Sodre, L., and Storrie-Lombardi, M.C.: 1995, {\it Monthly Notices of the Royal Astronomical Society} {\bf 275}, 567. doi:10.1093/mnras/275.3.567.


\bibitem[Nair and Abraham(2010)]{2010ApJS..186..427N}Nair, P.B. and Abraham, R.G.: 2010, {\it The Astrophysical Journal Supplement Series} {\bf 186}, 427. doi:10.1088/0067-0049/186/2/427.


\bibitem[Paturel \emph{et al.}(1989)]{1989A&AS...80..299P}Paturel, G., Fouque, P., Bottinelli, L., and Gouguenheim, L.: 1989, {\it Astronomy and Astrophysics Supplement Series} {\bf 80}, 299.


\bibitem[Ralph \emph{et al.}(2019)]{2019PASP..131j8011R}Ralph, N.O., Norris, R.P., Fang, G., Park, L.A.F., Galvin, T.J., Alger, M.J., and, ...: 2019, {\it Publications of the Astronomical Society of the Pacific} {\bf 131}, 108011. doi:10.1088/1538-3873/ab213d.

\bibitem[Schawinski \emph{et al.}(2014)]{2014MNRAS.440..889S}Schawinski, K., Urry, C.M., Simmons, B.D., Fortson, L., Kaviraj, S., Keel, W.C., and, ...: 2014, {\it Monthly Notices of the Royal Astronomical Society} {\bf 440}, 889. doi:10.1093/mnras/stu327.


\bibitem[Simmons \emph{et al.}(2017)]{2017MNRAS.464.4420S}Simmons, B.D., Lintott, C., Willett, K.W., Masters, K.L., Kartaltepe, J.S., H{\"a}u{\ss}ler, B., and, ...: 2017, {\it Monthly Notices of the Royal Astronomical Society} {\bf 464}, 4420. doi:10.1093/mnras/stw2587.

\bibitem[Stahl \emph{et al.}(2020)]{2020MNRAS.496.3553S}Stahl, B.E., Mart{\'\i}nez-Palomera, J., Zheng, W., de Jaeger, T., Filippenko, A.V., and Bloom, J.S.: 2020, {\it Monthly Notices of the Royal Astronomical Society} {\bf 496}, 3553. doi:10.1093/mnras/staa1706.

\bibitem[Storrie-Lombardi \emph{et al.}(1992)]{1992AAS...181.6508S}Storrie-Lombardi, M.C., Lahav, O., Sodre, L., and Storrie-Lombardi, L.: 1992, {\it American Astronomical Society Meeting Abstracts}.


\bibitem[VanderPlas \emph{et al.}(2012)]{2012cidu.conf...47V}VanderPlas, J., Connolly, A.J., Ivezic, Z., and Gray, A.: 2012, {\it Proceedings of Conference on Intelligent Data Understanding (CIDU}, 47. doi:10.1109/CIDU.2012.6382200.

\bibitem[Vander Plas, Connolly, and Ivezic(2014)]{2014AAS...22325301V}Vander Plas, J., Connolly, A.J., and Ivezic, Z.: 2014, {\it American Astronomical Society Meeting Abstracts \#223}.


\bibitem[Vasylenko \emph{et al.}(2019)]{2019OAP....32...46V}Vasylenko, M.Y., Dobrycheva, D.V., Vavilova, I.B., Melnyk, O.V., and Elyiv, A.A.: 2019, {\it Odessa Astronomical Publications} {\bf 32}, 46. doi:10.18524/1810-4215.2019.32.182538.

\bibitem[Vavilova, Melnyk, and Elyiv(2009)]{2009AN....330.1004V}Vavilova, I.B., Melnyk, O.V., and Elyiv, A.A.: 2009, {\it Astronomische Nachrichten} {\bf 330}, 1004. doi:10.1002/asna.200911281.


\bibitem[Vavilova \emph{et al.}(2015)]{2015KosNT..21c..94V}Vavilova, I.B., Ivashchenko, G.Y., Babyk, I.V., Sergijenko, O., Dobrycheva, D.V., Torbaniuk, O.O., and, ...: 2015, {\it Kosmichna Nauka i Tekhnologiya} {\bf 21}, 94. doi:10.15407/knit2015.05.094.


\bibitem[Vavilova, Elyiv, and Vasylenko(2018)]{2018RRPRA..23..244V}Vavilova, I.B., Elyiv, A.A., and Vasylenko, M.Y.: 2018, {\it Russian Radio Physics and Radio Astronomy} {\bf 23}, 244. doi:10.15407/rpra23.04.244.

\bibitem[Vavilova \emph{et al.}(2020)]{2020kdbd.book..307V}Vavilova, I., Dobrycheva, D., Vasylenko, M., Elyiv, A., and Melnyk, O.: 2020, {\it Knowledge Discovery in Big Data from Astronomy and Earth Observation}, 307. doi:10.1016/B978-0-12-819154-5.00028-X.


\bibitem[Wagner, Melnik, and Rucker(2019)]{2019EPSC...13..751W}Wagner, S., Melnik, V., and Rucker, H.: 2019, {\it EPSC-DPS Joint Meeting 2019}.

\bibitem[Wang, Cao, and Yu(2019)]{2019arXiv190604928W}Wang, S., Cao, J., and Yu, P.S.: 2019, {\it arXiv e-prints}, arXiv:1906.04928.

\bibitem[Way \emph{et al.}(2012)]{2012amld.book.....W}Way, M.J., Scargle, J.D., Ali, K.M., and Srivastava, A.N.: 2012, {\it Advances in Machine Learning and Data Mining for Astronomy, CRC Press, Taylor \& Francis Group, Eds.: Michael J. Way, Jeffrey D. Scargle, Kamal M. Ali, Ashok N. Srivastava}.


\bibitem[Willett \emph{et al.}(2013)]{2013MNRAS.435.2835W}Willett, K.W., Lintott, C.J., Bamford, S.P., Masters, K.L., Simmons, B.D., Casteels, K.R.V., and, ...: 2013, {\it Monthly Notices of the Royal Astronomical Society} {\bf 435}, 2835. doi:10.1093/mnras/stt1458.


\bibitem[York \emph{et al.}(2000)]{2000AJ....120.1579Y}York, D.G., Adelman, J., Anderson, J.E., Anderson, S.F., Annis, J., Bahcall, N.A., and, ...: 2000, {\it The Astronomical Journal} {\bf 120}, 1579. doi:10.1086/301513.



\end{thebibliography}

\end{document}